\documentclass[akbc,twoside,11pt,lettersize]{article}
\usepackage{akbc}

\usepackage{amsmath}
\usepackage{amssymb}
\usepackage{mathtools}
\usepackage{amsthm}
\usepackage{amsfonts}

\usepackage{microtype}
\usepackage{subfigure}
\usepackage{graphicx}
\usepackage{adjustbox}
\usepackage{xcolor, colortbl}
\usepackage{booktabs} 

\usepackage{hyperref}

\DeclareMathOperator*{\diag}{\mathrm{diag}}

\DeclareMathOperator*{\sigmoid}{\mathrm{\sigma}}

\DeclareMathOperator*{\logit}{\mathrm{logit}}

\akbcheading
\ShortHeadings{Paper Title}{Author list in the format of LastName1, LastName2, ... \& LastLastName}
\finalcopy % Uncomment for camera-ready version, but NOT for submission.
\begin{document}

\title{Pseudo-Riemannian Embedding Models for Multi-Relational Graph Representations}

\author{\name Saee Paliwal \email saee.paliwal@benevolent.ai \\
      \addr BenevolentAI \\
      London, United Kingdom
       \AND
       \name Angus Brayne \email angus.brayne@benevolent.ai \\
      \addr BenevolentAI\\
      London, United Kingdom
       \AND
       \name Benedek Fabian \email benedek.fabian@benevolent.ai \\
      \addr BenevolentAI\\
      London, United Kingdom
       \AND
       \name Maciej Wiatrak \email maciej.wiatrak@benevolent.ai \\
      \addr BenevolentAI\\
      London, United Kingdom
       \AND
       \name Aaron Sim \email aaron.sim@benevolent.ai \\
       \addr BenevolentAI\\
       London, United Kingdom}

% For research notes, remove the comment character in the line below.
% \researchnote

\maketitle
\begin{abstract}
% Embedding graph vertices in pseudo-Riemannian manifolds enables the incorporation of non-metric inductive biases into graph representation learning. 
In this paper we generalize single-relation pseudo-Riemannian graph embedding models to multi-relational networks, and show that the typical approach of encoding relations as manifold transformations translates from the Riemannian to the pseudo-Riemannian case. In addition we construct a view of relations as separate spacetime submanifolds of multi-time manifolds, and consider an interpolation between a pseudo-Riemannian embedding model and its Wick-rotated Riemannian counterpart. We validate these extensions in the task of link prediction, focusing on flat Lorentzian manifolds, and demonstrate their use in both knowledge graph completion and knowledge discovery in a biological domain.
\end{abstract}

% TLDR: We present PseudoE, a multi-relational extension of pseudo-Riemannian graph embedding models, and explore its application to the task of link prediction.
% Keywords: Machine Learning, Embeddings, Riemannian manifolds, Link prediction

%%%%%%%%%%%%%%%%%%%%%%%%%%%%%%%%%%%%%%%%%%%%%%%%%%%%%%%%%%%%%%%%%%%%%%%%%%%%%%%%%%%%%%%%%
\section{Introduction}
\label{Introduction}
A knowledge graph is a succinct abstraction of facts as a set of \textit{entities} and their pairwise \textit{relations}. Over the years, our ability to amass granular, heterogeneous relational data has increased substantially, allowing for graphical representations of entire systems, such as biology \cite{himmelstein2015heterogeneous} or society \cite{freebase-full}. Learning expressive representations of these graphs is an important first step in many machine learning-enabled applications, from recommender systems to drug discovery.

One effective class of methods for learning these whole-graph representations is node embedding models \cite{nickel2015review}. These models scale easily to large networks \citep{dettmers2018convolutional} and readily allow for applications such as node classification, clustering, and link prediction. The desire for low-dimensional yet expressive representations \cite{seshadhri2020impossibility} of nodes in these complex networks has prompted the exploration of more general Riemannian manifold embeddings as an effective means of incorporating useful inductive biases (see for example \citet{trouillon2016, nickel2017, gu2018learning, suzuki2019, lopez2021symmetric}).

A more recent development is pseudo-Riemannian manifold embeddings for graph representations \cite{sim2021directed}. Unlike their Riemannian manifold counterparts, pseudo-Riemannian embeddings are free from a host of metric space constraints, such as the upper bound on the number of disconnected nearest neighbors of a given node \cite{sun2015}. As identified in \citet{sim2021directed}, there are many real-world examples, where these constraints are routinely violated, and where the introduction of an indefinite metric enables a more faithful representation of those graph features.

There are, however, two shortcomings limiting the wider applicability of pseudo-Riemannian embedding models. First, they are restricted to single-relation graphs, and second, the lightcone structure of the loss function described in \citet{sim2021directed} is a very rigid constraint that may not be suitable for representing certain graphs with weak or few non-metric structures. For instance, many knowledge graphs, e.g. Freebase \cite{freebase-full}, contain up to $\sim 10^5$ relations, many of which are of the \textit{is-associated} or \textit{similar-to} types that can be well represented in metric spaces.

In this work, we propose \textbf{PseudoE}, a multi-relational extension of the pseudo-Riemannian embedding model from \citet{sim2021directed}. We model relations in two ways -- as endomorphisms, and as separate relation-specific spacetime submanifolds. In addition, to allow for increased representational flexibility, we introduce a set of node- and relation-specific bias terms to the training objective (similar to the node biases of \citet{balazevic2019}). We also consider a smooth interpolation between a pseudo-Riemannian embedding model and its Wick-rotated Riemannian counterpart, softening the constraints of the lightcone structure. For the purpose of this paper, we restrict ourselves to trivial flat pseudo-Riemannian manifolds, leaving the curved space generalisation to a future work. We validate PseudoE on a set of classic knowledge graph completion challenges and show that it is either competitive with or exceeds the state of the art. We also demonstrate the role of the bias terms in capturing graph structure and explore their application to gene prioritization for drug discovery.

%We choose to do this in order to most easily enable multi-time projections, and additionally, because the performance of flat pseudo-Riemannian manifolds can be competitive with their non-trivial counterparts \citet{sim2021directed}

%We illustrate the ability of PseudoE to variably capture directed versus symmetric edges by scaling the contribution of the pseudo-Riemannian and Wick-rotated Riemannian components of the probability function.
%%%%%%%%%%%%%%%%%%%%%%%%%%%%%%%%%%%%%%%%%%%%%%%%%%%%%%%%%%%%%%%%%%%%%%%%%%%%%%%%%%%%%%%%%
\section{Related Work}

Previous pseudo-Riemannian embedding models have employed three constant curvature spacetime manifolds -- Minkowski \cite{clough2017, sun2015}, anti-de Sitter \cite{sim2021directed} and de Sitter \cite{krioukov2012network} spacetimes, where the presence of the time dimension allows for the representation of directed graphs. However, these have only been applied to single-relation graphs with fixed node embeddings.

Spacetime coordinates have also been used in the Lorentzian model for hyperbolic embeddings \cite{nickel2018}. In this case, though, they are simply a parameterization of a (non-pseudo) Riemannian quadric surface, and also with fixed node embeddings.

Ultra-hyperbolic embeddings of \citet{law2020} introduced the idea of multiple-time manifolds. This work does not, however, consider the use of these additional time dimensions for representing multi-relational graphs, nor indeed prescribe any particular interpretation to the timelike dimensions, as is the case in \citet{sim2021directed} for single time dimensions.

The idea of relations as node transformations on vector spaces is well-established with many variants, ranging from translations \cite{bordes2013} to projections \citep{nickel2011, yang2015}. In \citet{balazevic2019}, a combination of separate transformations on general Riemannian geometries was proposed, generalizing both Euclidean and Poincar\'e embeddings to multi-relational graphs. Building on this, \citet{chami-etal-2020-low} variably compose relation-specific hyperbolic isometries using attention over these transformations. Here we further extend this body of work by applying these manifold maps to pseudo-Riemannian manifolds. These transformations, in combination with multi-time manifolds, allow us to extend the single-relation pseudo-Riemannian embedding model of \citet{sim2021directed} and explore its capacity to model multiple relations.

%%%%%%%%%%%%%%%%%%%%%%%%%%%%%%%%%%%%%%%%%%%%%%%%%%%%%%%%%%%%%%%%%%%%%%%%%%%%%%%%%%%%%%%%%
\section{Background} \label{background}
\subsection{Pseudo-Riemannian Embedding Models} \label{pse}
Let $G = (V, E)$ be a directed graph with $V = \{v_i\}_{i=1}^N$ the set of $N$ vertices (or nodes) and $E = \{(v_i, v_j)\}$ the set of directed edges represented by ordered node pairs, each containing a \textit{head}, $v_i$ and \textit{tail}, $v_j$ node. In node embedding models, each abstract node $v_i$ is mapped to a point $p_i$ on a manifold $\mathcal{M}$. $\mathcal{M}$ is in most cases endowed with the additional structure of a Riemannian metric $g$ -- a bilinear, symmetric, and positive-definite map on tangent vectors -- as this allows for unique geodesics to be defined between any two points on $\mathcal{M}$. This is useful, as the notion of node similarity will then have an unambiguous geometric counterpart in terms of geodesic distance.

%; in fact node embedding models are typically trained by optimising the alignment of some \textit{a priori} belief on node similarities (the edges themselves, correlation statistics, etc) with the set of node proximities in embedding space.

The geodesic uniqueness guarantee extends beyond Riemannian manifolds to the larger class of \textit{pseudo-Riemannian manifolds}, where $g$ is non-degenerate but no longer constrained to be positive definite. The metric has \textit{signature} $(n_t, n_x)$ where $n_t, n_x \in \mathbb{N}$ are the number of negative and positive eigenvalues, respectively, and $n \equiv n_t + n_x$ is the embedding dimension. The simplest example is the flat Minkowski spacetime manifold with $n_t=1$ and diagonal metric $g = \diag(-1, 1, \dotsc, 1)$. Given the coordinates $(x_0, \mathbf{x})$ and $(y_0, \mathbf{y})$ of two points $p$ and $q$ respectively, with $x_0, y_0$ the `time' and $\mathbf{x}, \mathbf{y} \in \mathbb{R}^{n_x}$ the `space' coordinates, the squared geodesic distance $s^2$ between $p$ and $q$ is
\begin{equation}
    s^2 = -(x_0 - y_0)^2 + \lvert\mathbf{x} - \mathbf{y}\rvert^2.
\end{equation}
Node embeddings on pseudo-Riemannian manifolds were first introduced in \citet{sun2015} and subsequently built upon in \citet{law2020} and  \citet{sim2021directed} for specific classes of manifolds, including ones with compact dimensions. 

Here we interpret \textit{Wick rotation} to be the process of converting a pseudo-Riemannian metric $g$ to a counterpart Riemannian metric $\tilde{g}$ by first choosing an orthogonal coordinate chart, e.g. via Gram-Schmidt orthogonalization, and then taking the absolute values of the resulting diagonal metric \citep{gao2018, visser2017}, i.e.,
\begin{equation} \label{wick}
    g \rightarrow \diag(a_0, \dotsc, a_{n-1}) \rightarrow \diag(|a_0|, \dotsc, |a_{n-1}|) \equiv \tilde{g},
\end{equation}
where $a_0, \dotsc, a_{n-1} \in \mathbb{R}$. In general, there are many ways to carry out the first step, and hence there is no canonical Wick-rotated Riemannian metric. Nevertheless, in this paper we only consider diagonal metrics and hence the procedure is unique. This process is needed for optimizing functions on pseudo-Riemannian manifolds \cite{gao2018}, but in this work it is primarily employed in a regularization term in the training loss function (see Section \ref{debias}).

The probability of an edge can be given by a function of the squared geodesic distance between its defining node pair via the \textit{Fermi-Dirac} (FD) distribution function \citep{krioukov2010, nickel2017}
\begin{equation}\label{fd}
    F_{(\tau, u, \alpha)}(s^2) := \frac{1}{e^{(\alpha s^2 - u) / \tau} + 1},
\end{equation}
with $x\in\mathbb{R}$ and parameters $\tau, u \geq 0$ and $0 \leq \alpha \leq 1$. For spacetime manifolds \citet{sim2021directed} proposed the \textit{Triple Fermi-Dirac} (TFD) function $\mathcal{F}(p, q) := k (F_1F_2F_3)^{1/3}$ with
\begin{equation}\label{tfd}
F_1 := F_{(\tau_1, u, 1)}(s^2), \quad
    F_2 := F_{(\tau_2, 0, \alpha)}(-\Delta t), \quad
    F_3 := F_{(\tau_2, 0, \alpha')}(\Delta t), \quad
\end{equation}
% with scale $k > 0$ and
% \begin{equation} \label{F123}
%     F_1 := F_{(\tau_1, u, 1)}(s^2), \quad
%     F_2 := F_{(\tau_2, 0, \alpha)}(-\Delta t), \quad
%     F_3 := F_{(\tau_2, 0, \alpha')}(\Delta t), \quad
% \end{equation}
where $F_1$, $F_2$, and $F_3$ are three FD distribution terms and $k > 0$. $\tau_1, \tau_2, u$, $\alpha$, and $\alpha'$ are the parameters from the component FD terms eq. \eqref{fd}, and $\Delta t \equiv x_0 - y_0$ the displacement in the time coordinate. In the TFD function, $F_1$ defines a lightcone partition of the spacetime manifold relative to each node in $V$ that concentrates the higher probabilities to its causal past and future. As illustrated in the top left panel of Figure \ref{illustration}, the probability decays into the past and future directions. Enforcing $\alpha \neq \alpha'$ in $F_2$ and $F_3$ introduces a time asymmetry to represent directed edge probabilities and allows for modulation of the prevalence of transitive relations.

%%%%%%%%%%%%%%%%%%%%%%%%%%%%%%%%%%%%%%%%%%%%%%%%%%%%%%%%%%%%%
\subsection{Multi-Relational Link Prediction}

The graphs of the previous section can be generalized to incorporate multiple edge types, i.e. for a given set of edge types/relations, $R$, the set of head-tail tuples $\{(v_i, v_j)\}$ is replaced by the set of head-relation-tail triples $\{(v_i, r_k, v_j)\},$ where $r_k \in R$ and $k \in [1, n_r] \equiv |R|$, and $n_r$ is the number of relations. These sets of observed triples constitute a knowledge graph, which we denote as $\hat{\mathcal{G}}$ and assume is noisy and incomplete relative to some underlying true set $\mathcal{G}$.

The task of link prediction \citep{nickel2015review} is to learn a \textit{score function} $\phi: \mathcal{G} \rightarrow \mathbb{R}$, such that we have probabilities
\begin{equation} \label{linkpredscore}
  \mathrm{P}\bigl[(v_i, r_k, v_j)\bigr] \equiv \sigmoid(\phi(v_i, r_k, v_j)),
\end{equation} 
over all possible triples, where $\sigmoid$ is the sigmoid function.

In knowledge graph embedding models, $\phi$ is often just a simple linear function on node and relation embedding coordinate vectors. For example in DistMult \citep{yang2015}, we have 
\begin{equation}
    \phi_\mathrm{D}(v_i, r_k, v_j) = \sum_{a := 1}^n(x_{p_i})_a (x_{r_k})_a (x_{p_j})_a,
\end{equation}
where $x_{p_*} \in \mathbb{R}^n$ is the coordinate vector of $p_*\in \mathcal{M} \equiv \mathbb{R}^n$, the node embedding of $v_*$. And in TransE \citep{bordes2013}, %\footnote{Assume an $L_1$-norm dissimilarity function}
\begin{equation}
    \phi_\mathrm{T}(v_i, r_k, v_k) := |x_{p_i} + x_{r_k} - x_{p_j}|.
\end{equation}

%%%%%%%%%%%%%%%%%%%%%%%%%%%%%%%%%%%%%%%%%%%%%%%%%%%%%%%%%%%%%%%%%%%%%%%%%%%%%%%%%%%%%%%%%
\section{PseudoE} \label{pseudoE} % Maybe methods instead?
\begin{figure}[ht!]
\begin{center}
\centerline{\includegraphics[width=\columnwidth]{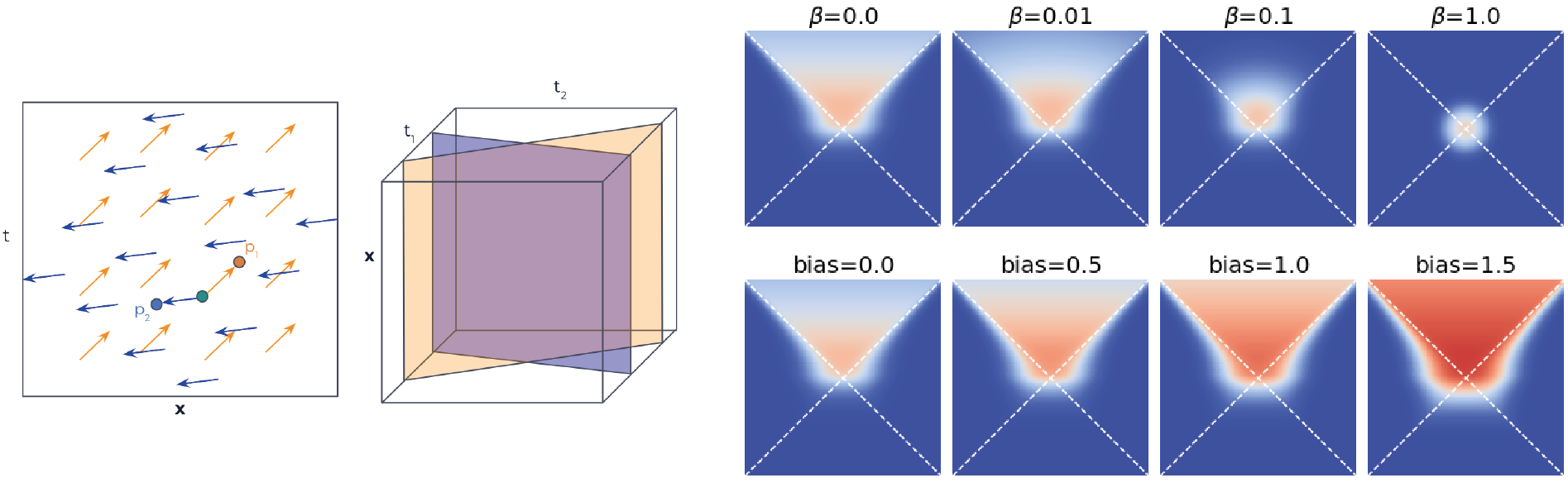}}
\caption{\textbf{Left/Middle}: Relations as diffeomorphisms induced by vector fields, and as spacetime submanifolds of a multi-time manifold. \textbf{Right top}: Interpolated likelihood $\mathcal{F}^{(\beta)}$ \eqref{regeq}; \textbf{Right bottom}: Bias effect $\sigma(\logit(\mathcal{F}) + \mathrm{bias})$ \eqref{phicombined}. The vertical axis is the timelike dimension for $\mathcal{M} = \mathbb{R}^{1,1}$.}
\label{illustration}
\end{center}
\vskip -0.2in
\end{figure}

We incorporate multiple relations via a combination of two strategies: 1. Relations as \textbf{endomorphisms}, and 2. relations as \textbf{spacetime submanifolds} of a node embedding manifold with multiple time dimensions. While the latter is specific to pseudo-Riemannian manifolds, the first is widely applied in Riemannian manifold embedding models. Node transformations are just point realizations of endomorphisms and is the method taken by all linear tensor factorization models and embedding models MuRE/MuRP \citet{balazevic2019}. Here we translate the implementation of the latter to pseudo-Riemannian manifolds and empirically validate their applicability when used together with the TFD function \eqref{tfd} to determine the probabilities in \eqref{linkpredscore}. Details of these two strategies are given below with an illustration in Figure \ref{illustration} (left/middle). We also introduce a Wick-rotated regularising Riemannian term and node and relation-specific bias terms, which we describe in more detail in Section \ref{debias}. We will refer to this combined model as PseudoE.
% \begin{figure}[ht!]
% \begin{center}
% \centerline{\includegraphics[width=\columnwidth]{figures/example_illustration.png}}
% \caption{\textbf{Left}: Example of relations as diffeomorphisms induced by vector fields, where each relation maps points along the integral curve solutions of the vector field; \textbf{Right}: Relations as spacetime submanifolds of a multi-time manifold.}
% \label{illustration}
% \end{center}
% \vskip -0.2in
% \end{figure}

%%%%%%%%%%%%%%%%%%%%%%%%%%%%%%%%%%%%%%%%%%%%%%%%%%%%%%%%%%%%%
\subsection{Relation Embeddings 1: Endomorphisms} \label{endomorphism}
The first strategy is to encode relations as node transformations, resulting in \textit{relation-specific node embeddings}. The framework developed in \citet{balazevic2019} for Riemannian manifolds is to map each relation to a pair of endomorphisms of $\mathcal{M}$, i.e. 
\begin{equation}\label{endoeq}
    r_k \, \mapsto \, (f_k, g_k),
\end{equation}
where $f_k, g_k: \mathcal{M}\rightarrow\mathcal{M}$ are the node transformations of the respective head and tail nodes in each triple candidate. Adapting the setup for our pseudo-Riemannian case, we replace the squared distance functions with our TFD function \eqref{tfd}.

%%%%%%%%%%%%%%%%%%%%%%%%%%%%%%%%%%%%%%%%%%%%%%%%%%%%%%%%%%%%%
\subsubsection{MuRE/P for pseudo-Riemannian manifolds}

One candidate for the score function $\phi$ in \eqref{linkpredscore} for a multi-relational knowledge graph given in terms of the single-relation TFD function \eqref{tfd} is
\begin{equation} \label{phi1}
    \phi_E(v_i, r_k, v_j) := \logit \bigl(\mathcal{F}(f_k(p_i), g_k(p_j))\bigr).
\end{equation}
where the subscript E refers to \textit{endomorphism}. This general definition allows a broad array of design options for $f_k$ and $g_k$, e.g. non-invertible, non-linear neural networks. In this paper we adapt the two transformation choices from \citet{balazevic2019} for Riemannian manifold embeddings as follows.

The first transformation is a TransE-like offset where each relation $r_k$ is mapped to a vector field $u_k$, which when restricted to any node embedding point $p\in \mathcal{M}$ gives the vector that defines the transformation. We have, for all $p\in\mathcal{M}$,
\begin{equation}\label{transelike}
    f_k(p) := \exp_{p}(u_k\rvert_{p}),
\end{equation}
where $\exp_p(*)$ is the exponential map on $\mathcal{M}$ at $p$. There are many ways to parameterize $u_k$ depending on the choice of $\mathcal{M}$. In this paper, we keep to the simplest example $\mathcal{M} = \mathbb{R}^{1, n-1}$ where eq. \eqref{transelike} is simply $f_k(p) = p + u_k\rvert_p$, for constant $u_k$.

The second transformation is DistMult-like component-wise scaling, where each relation $r_k$ is mapped to a diagonal $(1,1)$-tensor $R_k \in T_\mathbf{0}\mathcal{M} \otimes T_\mathbf{0}^*\mathcal{M}$ and 
\begin{equation}
    g_k(p) := \exp_\mathbf{0} (R_k \log_\mathbf{0}(p)),
\end{equation}
where $\log_\mathbf{0}: \mathcal{M}\rightarrow T_\mathbf{0}\mathcal{M}$ is the logarithm map. In the example $\mathcal{M} = \mathbb{R}^{1, n-1}$ we have $g_k(p) = R_k p$.

%%%%%%%%%%%%%%%%%%%%%%%%%%%%%%%%%%%%%%%%%%%%%%%%%%%%%%%%%%%%%
\subsection{Relation Embeddings 2: Spacetime submanifolds} \label{multitime}

In the previous section, the background geometry was fixed while the node embeddings were mapped to different points under relation-specific endomorphisms. In this section we take the dual view where node embeddings themselves remain fixed while the manifold itself adapts to different relations. 

The time displacement is explicitly used by the TFD function to provide a direction to the edges, so for multiple relations we can simply define a manifold with multiple time dimensions and introduce the procedure of a \textit{relation-specific time projection}. We note that this projection is a specific instance of an endomorphism, one, however, that is necessary for multi-time manifolds. Which is to say, any relation-specific endomorphism on a manifold with multiple time dimensions must define a submanifold with only a single time dimension.

% We choose to single out this instance because of the role the time dimension plays in calculating edge probabilities; essentially, any relation-specific endomorphism on a 

% one manifold with multiple time dimensions must define a submanifold with one single time dimension.

For a knowledge graph with $n_r$ relations, let $\mathcal{M}$ be a pseudo-Riemannian manifold with signature $(n_t, n_x)$, where $n_t > 1$. Next, for each relation $r_k$ we define a mapping $\tau_{k}: \mathcal{M} \rightarrow \mathcal{M}_k$ where $\mathcal{M}_k$ is a submanifold of $\mathcal{M}$ with signature $(1, n_x)$. Then the second candidate for the score function $\phi$ is
\begin{equation} \label{phi2}
    \phi_{SS}(v_i, r_k, v_j) := \logit\bigl(\mathcal{F}(\tau_k(p_i), \tau_k(p_j)\bigr).
\end{equation}
where the subscript SS refers to \textit{spacetime submanifold}. Just as for the endomorphisms \eqref{endoeq}, one has great freedom in designing $\tau_k$. In this work, we restrict ourselves to the simplest flat case of $\mathcal{M} = \mathbb{R}^{n_t, n_x}$ and for each relation $r_k$ we specify a vector $h_k \in \mathbb{R}^{n_t}$ such that the submanifold map is
\begin{equation}\label{flatproj}
    \tau_k: \bigl(\mathbf{t}, \mathbf{x}\bigr) \mapsto \bigl(t_k, \mathbf{x}\bigr) \equiv \bigl((h_k \cdot \mathbf{t}), \mathbf{x}\bigr),
\end{equation}
where $\mathbf{t}\in \mathbb{R}^{n_t}$ and $\mathbf{x}\in \mathbb{R}^{n_x}$ are, respectively, the time and space coordinates of a given node embedding. Following \citet{sim2021directed} (eq. 20) we also consider cylindrical Minkowski submanifolds by identifying $t_k \sim t_k + aC$, for some multiple of the circumference $C \in \mathbb{R}^+$.

In this paper we set $n_t < n_r$, imposing that the time coordinate parameters be shared between relations. In eq. \eqref{flatproj}, the submanifold time coordinate $t_k$ is a linear combination of the multiple time coordinates of the original manifold.
% This is similar to the approach of TuckER \cite{balazevic-etal-2019-tucker} where each relation matrix is a linear combination of a relatively small number of matrix slices from a shared tensor. 
%%%%%%%%%%%%%%%%%%%%%%%%%%%%%%%%%%%%%%%%%%%%%%%%%%%%%%%%%%%%%
\subsection{Wick-rotation regularisation and biases} \label{debias}

The lightcone structure of the TFD function \eqref{tfd} may turn out to be too strong of a constraint for certain graph structures (e.g. random geometric graphs) with relations that suit standard Riemannian backgrounds. To accommodate this we consider an interpolation between Riemannian and pseudo-Riemannian structures as follows. For a fixed mixing coefficient $0 \leq \beta \leq 1$, we replace the TFD function $\mathcal{F}$ with the weighted geometric mean
\begin{equation} \label{regeq}
    \mathcal{F}^{(\beta)} := \mathcal{F}^{1-\beta} \widetilde{F}^{\beta},
\end{equation}
where $\widetilde{F}\equiv F_1(\tilde{s}^2)$ is the Fermi-Dirac term from \eqref{tfd} with the squared geodesic distance $\tilde{s}^2$ calculated in the Wick-rotated space with metric $\tilde{g}$ \eqref{wick}. In Figure \ref{illustration} (top right), we plot an illustrative examples of the effect of the mixing; we observe a smooth interpolation between a lightcone structure with sharp boundaries and a Gaussian-like isotropic profile. 
% \begin{figure}[ht!]
% \vskip 0.2in
% \begin{center}
% \centerline{\includegraphics[width=\columnwidth]{figures/heatmap2.png}}
% \caption{\textbf{Top}: Interpolated likelihood $\mathcal{F}^{(\beta)}$ \eqref{regeq}; \textbf{Bottom}: Bias effect $\sigma(\logit(\mathcal{F}) + \mathrm{bias})$ \eqref{phicombined}. The vertical axis is the timelike dimension for $\mathcal{M} = \mathbb{R}^{1,1}$.}
% \label{lossplot}
% \end{center}
% \vskip -0.2in
% \end{figure}

The final addition to our model is a set of specific bias terms for the nodes -- $b_i, b_j$, and relations -- $c_k$. The node biases introduced in \citet{balazevic2019} for Riemannian manifold embeddings were interpreted as defining the size of spheres of influences; in the pseudo-Riemannian case, as shown in Figure \ref{illustration} (bottom right), the geometrical picture is more akin to \emph{cones} of influence, where the time/spacelike partition of the manifold, and hence the pseudo-Riemannian embedding model as a whole, is preserved. 

Combining the elements of our model -- merging $\phi_E$ (eq. \eqref{phi1}) and $\phi_{SS}$ (eq. \eqref{phi2}) and including the features in eqs.\eqref{flatproj}, and \eqref{regeq} -- we define our PseudoE score function as
\begin{equation}
\label{phicombined}
    \phi(v_i, r_k, v_j) = \logit \Bigl(\mathcal{F^{(\beta)}}\bigl[(f_k\circ\tau_k)(p_i), (g_k\circ\tau_k)(p_j)\bigr]\Bigr)
    + b_i + b_j + c_k,
\end{equation}
where the manifold map is performed after the spacetime projection.

%%%%%%%%%%%%%%%%%%%%%%%%%%%%%%%%%%%%%%%%%%%%%%%%%%%%%%%%%%%%%%%%%%%%%%%%%%%%%%%%%%%%%%%%%
\section{Results}
\label{results}
In this section we evaluate PseudoE \eqref{phicombined} on a standard set of link prediction tasks and demonstrate how its separate components capture different aspects of graph data. For simplicity, we restrict ourselves to the trivial flat pseudo-Riemannian manifolds with metric $g = \diag(-1, \dotsc, -1, 1, \dotsc, 1)$, leaving the study of curved spaces to a later work.

%%%%%%%%%%%%%%%%%%%%%%%%%%%%%%%%%%%%%%%%%%%%%%%%%%%%%%%%%%%%%
\subsection{Experimental Setup} 

\subsubsection{Datasets}\label{Datasets}
We run our experiments on the two standard link prediction datasets: FB15K-237 and WN18RR, as well as the Hetionet scientific knowledge graph \cite{himmelstein2015heterogeneous}. A full description is provided in Appendix \ref{dataset_stats}.

\subsubsection{Training}
For FB15K-237 and WN18RR, we use the common data augmentation technique \cite{dettmers2018convolutional} of adding reversed triples, i.e. for every $(v_i,r_k,v_j)$ triple that exists, we append $(v_j, \tilde{r}_k, v_i)$. For Hetionet, we follow \citet{nadkarni2021} and omit this step.

The model is trained by minimizing the negative log-likelihood loss
\begin{equation}\label{nll}
\small
\begin{split}
    \mathcal{L} = -\sum_{\substack{i,j \in V, k \in R \\ (v_i, r_k, v_j) \in \hat{\mathcal{G}}}}\biggl[\log \sigma(\phi(v_i, r_k, v_j)) +
     \sum_{\substack{a=1 \\ l_a\in V}}^{m / 2}\log (1- \sigma(\phi(v_i, r_k, v_{l_a}))) + \sum_{\substack{b=1 \\ l_b\in V}}^{m / 2}\log (1 - \sigma(\phi(v_{l_b}, r_k, v_j))) \biggr],
\end{split}
\end{equation}
where $m$ is the even number of random negative node samples ($v_{l_a}, v_{l_b}$) per triple. In the case where data augmentation is performed, we drop the last summation term in \eqref{nll} and perform all $m$ negative replacements solely on the tail node. We train with minibatch stochastic gradient descent and experiment with both Adam \cite{kingma2014adam} and SM3 \cite{anil2019memory} optimizers. All node and relation embeddings are randomly initialised with the normal distribution $N(0, \sigma_i^2)$. We perform early stopping on the evaluation metric (see next section).

We perform a round of hyperparameter tuning via random search over the validation set evaluation metric. The optimal hyperparameters selected from each dataset and model combination are given in Table \ref{hyperparameters}.

\subsubsection{Evaluation}
%As is the convention for link prediction evaluations, 
We evaluate our models using the mean reciprocal rank (MRR) and the hits@k information retrieval metrics \cite{schutze2008introduction}. Following convention, we use the \textit{filtered} version of the metric \cite{bordes2013} where each test set triple is ranked against all others obtained by swapping out the tail node, but excluding those triples that can be found in the full dataset $\hat{\mathcal{G}}$. 
In the case of Hetionet, in order to compare with the results presented in \citet{nadkarni2021}, we use their predefined fixed-sized ($N=80$) set of negative samples of matching node type for each head-relation pair for both validation and testing.

%%%%%%%%%%%%%%%%%%%%%%%%%%%%%%%%%%%%%%%%%%%%%%%%%%%%%%%%%%%%%
\subsection{Link prediction performance}

\begin{table*}[t!]
\centering
\begin{adjustbox}{max width=0.9\textwidth}
\begin{tabular}{lccccccccc}
\toprule
& \multicolumn{2}{c}{WN18RR} &  & \multicolumn{2}{c}{FB15K-237} & & \multicolumn{2}{c}{Hetionet (small)}\\ 
\cline{2-3}\cline{5-6} \cline{8-9} 
& MRR & Hits@10 &  & MRR & Hits@10 & & MRR & Hits@10 \\ 
\toprule
TransE \citep{bordes2013} & 0.226 & 0.501 & & 0.294 & 0.465 & & 0.502 & 0.798 \\  
DistMult \citep{yang2015} & 0.430 & 0.490 & & 0.241 & 0.419 & & 0.460 & 0.778 \\ 
ComplEx \citep{trouillon2016} & 0.440 & 0.510 & & 0.247 & 0.428 & & 0.459 & 0.778 \\ 
Rescal \citep{wang-rescal} & 0.420 & 0.447 & & 0.270 & 0.427 & & -- & -- \\
TuckER \citep{balazevic-etal-2019-tucker} & 0.470 & 0.526 & & \textbf{0.358} & \textbf{0.544} & & -- & --\\
MuRE \cite{balazevic2019}  & 0.475 & 0.554 & & 0.336 & 0.521 & & 0.527* & 0.809* \\
%RotE \cite{chami-etal-2020-low} & 0.494 & 0.585 & & 0.346 & 0.538 & & -- & -- \\
RotE* \cite{chami-etal-2020-low} & 0.494 & 0.585 & & 0.346 & 0.538 & & -- & -- \\
AttE* \cite{chami-etal-2020-low} & 0.490 & 0.581 & & 0.351 & 0.543 & & -- & -- \\
\midrule 
MuRP \cite{balazevic2019}  & 0.481 & 0.566 & & 0.335 & 0.518 & &  -- & -- \\
%RotH \cite{chami-etal-2020-low} & \textbf{0.496} & \textbf{0.586} & & 0.344 & 0.535 & & -- & -- \\
RotH* \cite{chami-etal-2020-low} & \textbf{0.496} & \textbf{0.586} & & 0.344 & 0.535 & & -- & --  \\
AttH* \cite{chami-etal-2020-low} & 0.486 & 0.573 & & 0.348 & 0.568 & & -- & --  \\
\midrule
%PseudoE: DT, no bias, no MT & TBD & TBD & & 0.307 & 0.475 &  & \cellcolor{lightgray}\textbf{0.546} & \cellcolor{lightgray}\textbf{0.815} \\ % Runs: Hetionet:  hetionet-2-ecoeff 
PseudoE (MT only) &  0.314 & 0.433 & & 0.273 & 0.451 & & 0.543 & 0.812 \\ % Wordnet: sgd-sync-8a3, FB15K: sgd-sync-7a2,  Hetionet: hetionet-2-only-mt
PseudoE (DT only) & \cellcolor{lightgray}0.474 & \cellcolor{lightgray}0.567 & & \cellcolor{lightgray}0.351 & \cellcolor{lightgray}0.539 & & 0.538 & 0.813 \\  % Runs: Wordnet: sgd-sync-10, FB15K: sgd-sync-9a5, Hetionet: hetionet-2-ecoeff-withbias2 
PseudoE (both) & 0.473 & 0.565 & & 0.351 & 0.536 &  & \cellcolor{lightgray}\textbf{0.544}& \cellcolor{lightgray}\textbf{0.813} \\  % Runs: Wordnet: sgd-sync-8a6, FB15K: sgd-sync-7a8, Hetionet: heitonet-2a10
\bottomrule
\end{tabular}
\end{adjustbox}
 \caption{Link prediction results. Best performance by metric is in \textbf{bold}, and the highest performing PseudoE variant is shaded in gray. Hetionet baselines are taken from \citet{nadkarni2021} with the exception of MuRE which is PseudoE run with $\beta=1$. DT: DistMult-TransE scoring function, MT: multi-time relations. *Shown here are the best results by dataset of the RotE/H, RefE/H and AttE/H variants \citet{chami-etal-2020-low}.}
\label{sota_results}
\end{table*}
We run our series of link prediction experiments testing three separate PseudoE variants -- one with the DistMult-TransE endomorphism only (DT), one with submanifolds of multi-time manifolds only (MT), and one with both implementations. We compare these to several common tensor factorization baselines, as well as the MuRE/P and AttE/H embedding models. The results are shown in Table \ref{sota_results}.
% Because PseudoE is defined on flat Lorentzian manifolds, the relevant baselines are the Euclidean-based models, which excludes MuRP and AttH. The results of these hyperbolic models are included as a point of reference, particularly because MuRP forms the basis of the design of PseudoE.

PseudoE is highly competitive on all three datasets, coming in either top by a good margin (Hetionet) or 2nd/3rd (FB15K-237 and WN18RR) among the comparable flat-space and linear models. For FB15K-237 specifically, it was first speculated in \citet{balazevic2019} that the relatively large number of relations benefit from the parameter sharing feature of the relational embeddings in TuckER, the most performant model. Therefore, the close performance of PseudoE demonstrates its ability to model a large number of relations without either explicit parameter sharing or, in our implementation, curved geometries. 
 
Next, it is clear from the model ablation results comparing MT, DT, and both model components together that the best combination of methods for encoding relations is highly dependent on the dataset used. For instance, the spacetime submanifold model performs poorly on both WN18RR and FB15K-237, but for Hetionet, it outperforms the endomorphism approach, with the combination of the two being the optimal setup. An obvious extension of this work would be to explore more expressive multi-time projections to explore the broader impact of this model feature on datasets beyond Hetionet.

Finally, we note that the single-relation pseudo-Riemannian baseline in \citet{sim2021directed} cannot do better than an assignment of relation types according to the frequencies observed in the training data. Even assuming perfect classification of edges vs. non-edges, the performances on both WN18RR and FB15K-237 datasets would be close to random.

% Unsurprisingly, MuRP, which employs hyperbolic graph embeddings that are ideal for modeling hierarchies, continues to be the optimal model for WN18RR, as this dataset contains relations arranged in a highly hierarchical structure. Notably, PseudoE with a Distmult-Transe scoring function and bias terms, and a mixing coefficient of 0.18, performs competitively with MuRE, the second most performant method on this dataset. For FB15K-237, the same PseudoE model variant performs competitively with the most performant model, TuckER, a bilinear model on a Euclidean manifold. This demonstrates the flexibility of the PseudoE variants to model hierarchical and undirected networks. Finally, PseudoE outperforms other methods on the Hetionet (small) dataset. This is expected, as PseudoE is designed to capture directed graphs and Hetionet's relations are directed, save for \textit{associated with}.

%%%%%%%%%%%%%%%%%%%%%%%%%%%%%%%%%%%%%%%%%%%%%%%%%%%%%%%%%%%%%
\subsection{Interpolating between geometries with $\beta$}
% \begin{itemize}
%     \item Hetionet experiment, varying euclidean coefficient
%     \item Qualitative Hetionet experiment, bias terms
%     \item Triangle-rich spare network recovery
% \end{itemize}
\begin{figure}[h!] 
  \centering
  \includegraphics[width=0.5\columnwidth]{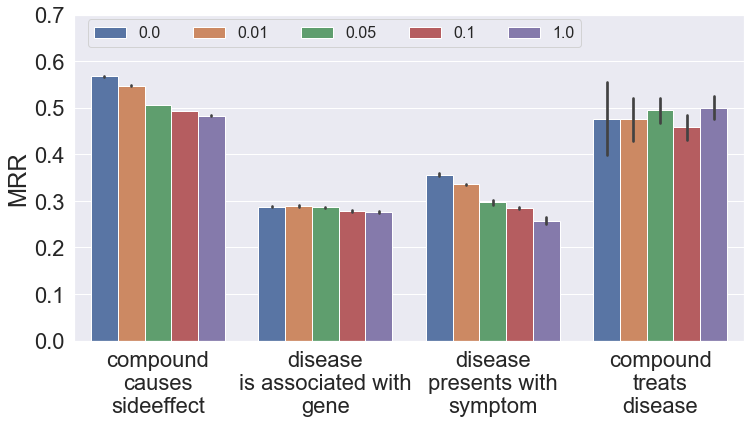}
  \caption{MRR by relation, varying the mixing coefficient from $\beta =0$ (Minkowski) to $\beta=1$ (Euclidean), on Hetionet. Error bars represent $\pm 1$ s.d. across four runs.}
\label{euclidean_coeff}
\end{figure}

In this section, we explore the effect of varying the weight coefficient $\beta$ \eqref{regeq} on link prediction performance. We found that the effect of the mixing parameter, assessed on an initial run of the DT model using smaller embedding size (d=200), was positive. For WN18RR, we achieve an MRR=0.472 ($\beta>$0) vs. 0.445 ($\beta$=0, i.e. no mixing). FB15-237 we achieve MRR=0.347 ($\beta>$0) vs. 0.318 ($\beta$=0). The highest performing PseudoE models on WN18RR, FB15K-237, and Hetionet in Table \ref{sota_results} had $\beta=0.18, \,0.30, \, \text{and} \, 0.03$ respectively. The relatively high value for FB15K-237 is not unsurprising given that it is a large heterogeneous knowledge graph with a large number of similarity-type relations for which a sharp lightcone likelihood profile may not be appropriate. 

We now observe how the relative contributions of a Minkowski geometry ($\beta=0$) vs. its Wick-rotated Euclidean ($\beta=1$) geometry affect the prediction of individual triples with the various directed or symmetric relations in Hetionet (Figure \ref{euclidean_coeff}). Notably, for the directed relations \textit{causes} and \textit{presents with}, we observe a clear drop in performance as the pseudo-Riemannian TFD likelihood function \eqref{tfd} gives way to an increasingly isotropic form (see Figure \ref{illustration} right). As the TFD function is designed specifically for such relations with non-metric properties \cite{sim2021directed}, it is a validation that the pseudo-Riemannian model outperforms on these specific relations. On the contrary, varying $\beta$ has a negligible impact on the performance of the undirected \textit{is associated with} relationship predictions. Finally, we note that the performance on the \textit{treats} relation is inconclusive, due to the small number of edges for this relation in the dataset ($N_\text{treats} = 597$ vs. $N_\text{causes}=110569$).

\begin{figure*}[ht] 
    \centering
  \includegraphics[width=0.92\textwidth]{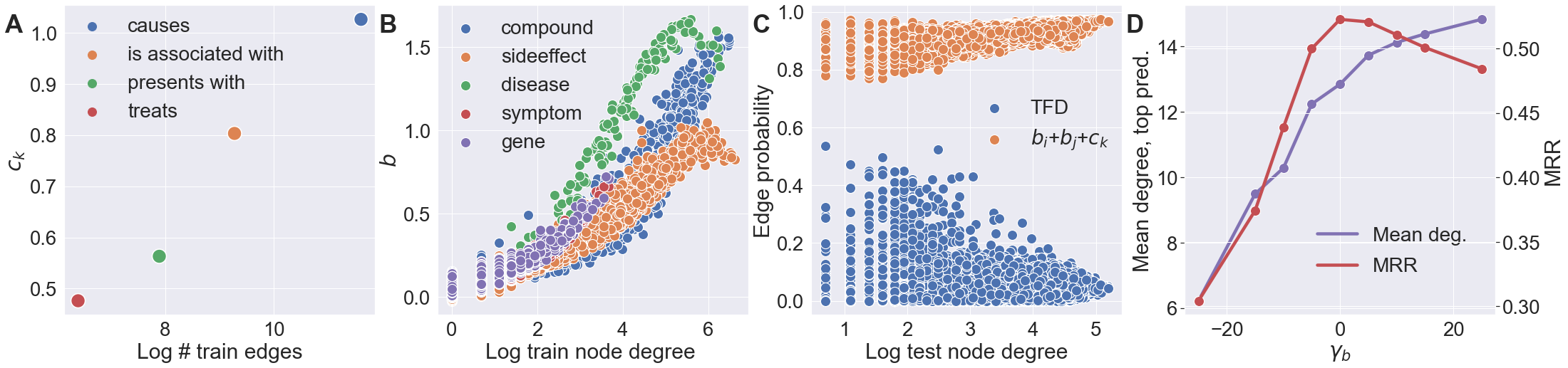}
  \caption{\textbf{A}: Relation bias terms $c_k$ vs. edge count. \textbf{B}: Node bias term $b$ vs. node degree. \textbf{C}: Components of the composite score -- the TFD component and the combined bias terms $b_i+b_j+c_k$ -- vs. edge probability. \textbf{D}: Average degree of top-ranking prediction, scaling the target bias via $\gamma_b$.}
  \label{debiasing_kg}
\end{figure*}

%%%%%%%%%%%%%%%%%%%%%%%%%%%%%%%%%%%%%%%%%%%%%%%%%%%%%%%%%%%%%
\subsection{Removing bias from KG link predictions} \label{debias}

We explore the role that $c_k$, $b_i$, and $b_j$ \eqref{phicombined} play in disentangling the dataset biases related to relation prevalence and node degrees from our predictions. On the model trained on Hetionet, we observe a positive log-linear relationship between $c_k$ and the relation prevalence (Figure \ref{debiasing_kg}A), and a similarly positive relationship between the combined node biases $b=b_i + b_j$ and node degree across all five node types (Figure \ref{debiasing_kg}B).  Importantly, Figure \ref{debiasing_kg}C shows a strong positive correlation (Pearson's r) between node degree and the node bias component of edge probability ($r(15566)=0.69$, $p<.001$), versus a negligible negative relationship ($r(15566)=-0.05$, $p<.001$) between the TFD component. This ability offers an opportunity to factor out the node degree-bias of link predictions in a principled way at test time. We explore the application of this scaling to gene prioritization in drug target identification in Appendix Section \ref{qualitative}.

\section{Summary}

In this paper we introduced PseudoE, an extension of pseudo-Riemannian embeddings to multiple relations. For link prediction, PseudoE is competitive on FB15K-237 and WN18RR, and is state of the art amongst linear models on a subset of the Hetionet dataset. This work can be extended to curved pseudo-Riemannian manifolds and applied to a broader set of applications such as node classification.

% \acks{...
% }

\bibliography{pseudoe}
\bibliographystyle{plainnat}
%%%%%%%%%%%%%%%%%%%%%%%%%%%%%%%%%%%%%%%%%%%%%%%%%%%%%%%%%%%%%%%%%%%%%%%%%%%%%%%%%%%%%%%%%
\appendix

\setcounter{table}{0}
\setcounter{figure}{0}
\renewcommand{\thetable}{A\arabic{table}}
\renewcommand{\thefigure}{A\arabic{figure}}

\section{Datasets}
\subsection{FB15K-237} FB15K \cite{bordes2013} is a subset of the Freebase dataset \cite{freebase-full}, a collection of triples representing general history facts. FB15K-237 \cite{toutanova-etal-2015-representing} is the subset of FB15K where, to minimize data leakage, all triples with trivial inverse relations of training relations are removed from the validation and test sets.

\subsection{WN18RR} WordNet \cite{miller1995wordnet} is an acyclic, hierarchical, tree-like network of nouns, each with relatively few ancestors and many descendants. WN18RR \cite{dettmers2018convolutional}, a subset of WordNet, is comprised of eleven relations, and, similar to FB15K-237, is cleaned for inverse relation test set leakage.

\subsection{Hetionet (small)} Hetionet is a biomedical dataset that integrates several publicly-available scientific databases, including the Unified Medical Language System (UMLS) \cite{bodenreider2004unified}, GeneOntology \cite{ashburner2000gene}, and DisGenNET \cite{pinero2016disgenet}. In order to facilitate comparison of performance metrics, we restrict Hetionet to the subset used in \citet{nadkarni2021} and \citet{alshahrani2021application} -- referred to here as ``Hetionet (small)" -- that includes four relations, (\textit{treats}, \textit{presents}, \textit{associates}, and \textit{causes}), linking five entity classes (\textit{compounds}, \textit{diseases}, \textit{genes}, \textit{side effects}, and \textit{symptoms}) in a directed acyclic graph.

\label{sec:dataset_stats}
\begin{table}[htb!]
\centering
\begin{tabular}{lccccc}
\toprule
& & & \multicolumn{3}{c}{\# Edges}\\
\cmidrule{4-6}
 & \# Ent. & \# Rel. & Train & Valid & Test \\
\midrule
FB15K-237 & 14,541 & 237 & 272,114 & 17,534 & 20,465 \\
WN18RR & 40,943 & 11 & 86,836 & 3,033 & 3,133 \\ 
Hetionet (small) & 12,733 & 4 & 124,543 & 15,566 & 15,567 \\
\bottomrule
\end{tabular}
\caption{Dataset specificiation for link prediction task.}
\label{dataset_stats}
\end{table}

\section{Training hyperparameters}
\label{sec:hyperparameters}

Hyperparameters below relate to the models in Table \ref{sota_results}. These hyperparameters were selected based on random search maximising the validation MRR.

\begin{table*}[h!]
\begin{adjustbox}{max width=0.999\textwidth}
\begin{tabular}{l @{\extracolsep{4pt}}ccccccccc@{}}
\toprule
Model                       &                                  \multicolumn{3}{c}{PseudoE (MT only)} &                                  \multicolumn{3}{c}{PseudoE (DT only)} &                                     \multicolumn{3}{c}{PseudoE (both)} \\
\cline{2-4} \cline{5-7} \cline{8-10} 
Dataset                     &                                            Wordnet &                                              FB15K &                                           Hetionet &                                            Wordnet &                                              FB15K &                                           Hetionet &                                            Wordnet &                                              FB15K &                                           Hetionet \\
\toprule
$\alpha$                       &                                             0.3673 &                                           0.136206 &                                            0.10124 &                                             0.3673 &                                           0.136206 &                                            0.10124 &                                             0.3673 &                                           0.136206 &                                            0.10124 \\
$\alpha'$                     &                                            0.75182 &                                           0.971685 &                                                1.0 &                                            0.75182 &                                           0.971685 &                                                1.0 &                                            0.75182 &                                           0.971685 &                                                1.0 \\
$u$                   &                                           0.040226 &                                            0.09592 &                                               0.03 &                                           0.040226 &                                            0.09592 &                                               0.03 &                                           0.040226 &                                            0.09592 &                                               0.03 \\
$\tau_1$                     &                                            0.29015 &                                           0.129815 &                                            0.11071 &                                            0.29015 &                                           0.129815 &                                            0.11071 &                                            0.29015 &                                           0.129815 &                                            0.11071 \\
$\tau_2$                    &                                            0.21697 &                                           0.086457 &                                            0.06277 &                                            0.21697 &                                           0.086457 &                                            0.06277 &                                            0.21697 &                                           0.086457 &                                            0.06277 \\
$n_x$               &                                                500 &                                                200 &                                                200 &                                                500 &                                                500 &                                                200 &                                                500 &                                                500 &                                                200 \\
$n_t - 1$              &                                                  40 &                                                 40 &                                                  1 &                                                  0 &                                                  0 &                                                  0 &                                                  1 &                                                  2 &                                                  1 \\
$\beta$             &                                                0.18 &                                                0.0 &                                                0.0 &                                               0.18 &                                                0.15 &                                                0.0 &                                               0.18 &                                                0.15 &                                                0.0 \\
circumference               &                                               - &                                               - &                                               - &                                               - &                                                8.0 &                                                8.0 &                                               - &                                               - &                                                6.0 \\
$\sigma_i$ init scale                  &                                              0.001 &                                              0.001 &                                            0.02255 &                                              0.001 &                                              0.001 &                                            0.02255 &                                              0.001 &                                              0.001 &                                            0.02255 \\
Learning rate                          &                                                0.08 &                                                0.1 &                                             0.0002 &                                                0.08 &                                             0.0001 &                                             0.0002 &                                             0.08 &                                             0.0001 &                                             0.0002 \\
Batch size                  &                                                128 &                                                128 &                                                100 &                                                128 &                                                128 &                                                100 &                                                128 &                                                128 &                                                100 \\
Negative samples $m$                &                                                 50 &                                                 50 &                                                 20 &                                                 50 &                                                 50 &                                                 20 &                                                 50 &                                                 50 &                                                 20 \\

optimizer                   &                                          SM3 &                                          SM3 &                                               Adam &                                          SM3 &                                               Adam &                                               Adam &                                               Adam &                                               Adam &                                               Adam \\
                                                1 \\
\hline
\end{tabular}
\end{adjustbox}
\caption{Optimal model and training hyperparameters.} % and $b_n$ and $b_r$ are constant coefficients added to $b_i + b_j$ and $c_k$ respectively.}
\label{hyperparameters}
\end{table*}
We observe that sensitivity to specific hyperparameters is highly dataset-specific, with WN18RR being the dataset that required the most careful tuning. However, we note that this sensitivity is not unique to PseudoE and we observe very similar optimization challenges when reproducing the results of MuRE/P. Also, we note that many of the parameters are common to many models -- for instance $\tau_1$ and $\tau_2$ are analogous to the softmax temperature and $u$ to a loss-margin radius -- or involved in parameterizing the geometry. There are, in effect, just two additional hyperparameters specific to PseudoE: the mixing parameter $\beta$, and $n_t$, the number of time dimensions.

\section{Qualitative assessment of gene precedence with varying bias} \label{qualitative}

The negligible correlation of the TFD term with node degree, discussed in Section \ref{debias} offers an opportunity to factor out the node degree-bias of link predictions in a principled way -- at test time, one simply scales the appropriate bias terms to up- or down-weight those predictions that are poorly- or well-connected respectively, i.e. 
\begin{equation}
b_i \rightarrow \gamma_b b_i,
\end{equation}
for some scaling factor $\gamma_b \in \mathbb{R}.$

In Figure \ref{debiasing_kg}D we see that tuning the target bias varies the mean node degree of the top gene prediction across all test-set diseases in Hetionet. This allows us to variably prioritize either those genes that are novel and poorly connected (low $\gamma_b$) or are well studied and widely connected (high $\gamma_b$), trading off model performance for prediction novelty. A real-world application for this capability is in the area of gene prioritization in commercial drug discovery, as biologists seek out genes that are plausibly implicated in a particular disease (high model performance) while being both safe to target and novel, i.e. not being involved in many biological processes, or already targeted for treatment (low node degree in the knowledge graph).

Taking breast cancer as an example, Table \ref{breast_cancer} shows the 10 top-ranking associated genes for three levels of $\gamma_{b}$. Through target triage, similar to the process described in \cite{paliwal2020preclinical}, we annotated those targets that were non-specific cancer genes, those targets specific to breast cancer, those targets that are interesting and understudied to date, and those targets that were irrelevant. What is immediately noteworthy about the $\gamma_{b}=25$ list is the prevalence of generic cancer genes, like TNF and TP53. While the $\gamma_{b}=1$ case -- the original trained model -- unearths disease-specific genes, such as RAD51, the $\gamma_{b}=-25$ case reveals a handful of genes that are understudied (i.e. low degree) yet interesting, like DTX3 or RPS6KB2, with plausible or preliminary existing links to breast cancer, making them of particular interest in the context of drug discovery.

\begin{table}[ht!]
\centering
\small
\begin{tabular}{cccc}
\toprule
Rank & $\gamma_{b}=25$ & $\gamma_{b}=1$ & $\gamma_{b}=-25$ \\
\midrule
1 & \cellcolor{lightgray}TNF & COL7A1 & DTX3 $\bigstar$ \\
2 & \cellcolor{lightgray}TP53 & RAD51 $\blacksquare$ & MRPS23 $\bigstar$ \\
3 & PTGS2 & SYNJ2 $\blacksquare$  & RPS6KB2 $\bigstar$ \\
4 & \cellcolor{lightgray}IL6 & SLC39A7 & ABRAXAS1 $\bigstar$ \\
5 & \cellcolor{lightgray} MMP9 & MRE11 $\blacksquare$  & \cellcolor{lightgray} CDA \\
6 & ALB $\times$ & \cellcolor{lightgray}FGF3 & MAP2K7 \\ 
7 & \cellcolor{lightgray} CXCL8  & SFRP1 $\blacksquare$  & PCDHGB6 $\times$ \\
8 & \cellcolor{lightgray}IL2 & UBE2C $\blacksquare$ & ESRRA $\times$ \\ 
9 & \cellcolor{lightgray}AKT1 & TENT5A $\times$ & RBM3 $\bigstar$ \\ 
10 & \cellcolor{lightgray}TGFB1 & BABAM1 $\blacksquare$ & \cellcolor{lightgray}S100A4 \\
\bottomrule
\end{tabular}
\caption{Gene prioritization for breast cancer (BC). $\gamma_b=1$ is the original model. Shaded cells: generic cancer genes. $\blacksquare$: genes well-studied and established as linked to BC, $\times$: genes irrelevant/uninteresting for BC, $\bigstar$: genes relevant, interesting, and surprising for BC therapy research.}
\label{breast_cancer}
\end{table}

% Gene Nodes
% RPS6KB2 only associated with 4.3\% of breast cancers, involvced in GBM stem cells so interesting for breast cancer (https://www.ncbi.nlm.nih.gov/pmc/articles/PMC7072743/). 
% ALB is obvious for liver cancer with possible but not super interesting relationship with breast cancer

% DTX3, MRPS23 - interesting, few studies
% RPS6KB2 - Indirect links, definite interesting 
% ABRAX1 - Few studies, relevant to breast cancer biology, interesting
% CDA - stem cells, GWAS hit
% MAP2K7 - not so novel, well studied family of proteins. Good target in cancer 
% PCDHGB6 - very low data, but hits relevant mechanisms in cancer & BrCA
% ESRRA - few studes but not interesting target, estrogen receptor agonist
% RBM3 - interesting, relevant, studied in different cancers]
%%%%%%%%%%%%%%%%%%%%%%%%%%%%%%%%%%%%%%%%%%%%%%%%%%%%%%%%%%%%%

\end{document}